\documentclass[lettersize,journal]{IEEEtran}
\usepackage{amsmath,amsfonts}
\usepackage{algorithmic}
\usepackage{array}
\usepackage[caption=false,font=scriptsize,labelfont=rm,textfont=rm]{subfig}
\usepackage{textcomp}
\usepackage{stfloats}
\usepackage{url}
\usepackage{verbatim}
\usepackage{graphicx}
\usepackage{makecell}
\usepackage{balance}

\usepackage{multirow}
\usepackage{cite}
\usepackage{cases}
\usepackage{amssymb}
\usepackage[ruled]{algorithm2e}
\usepackage{bm}
\usepackage{color,xcolor}

\usepackage{amsthm}

\allowdisplaybreaks[3]

\begin{document}
	
	\title{Dual-Waveguide Pinching Antennas for PLS: Parallel Placement or Orthogonal Placement?}
	\author{Yang Lu,~\IEEEmembership{Senior Member,~IEEE}, Xinke Xie, Yanqing Xu,~\IEEEmembership{Member,~IEEE}, Bo Ai,~\IEEEmembership{Fellow,~IEEE}, \\ Octavia A. Dobre,~\IEEEmembership{Fellow,~IEEE}, and Arumugam Nallanathan,~\IEEEmembership{Fellow,~IEEE}
    
    \thanks{Yang Lu and Xinke Xie are with the State Key Laboratory of Advanced Rail Autonomous Operation, and also with the School of Computer Science and Technology, Beijing Jiaotong University, Beijing 100044, China (e-mail: yanglu@bjtu.edu.cn, 25110159@bjtu.edu.cn).}
    \thanks{Yanqing Xu is with the School of Science and Engineering, The Chinese University of Hong Kong, Shenzhen, 518172, China (email: xuyanqing@cuhk.edu.cn).}
    \thanks{Bo Ai is with the School of Electronics and Information Engineering, Beijing Jiaotong University, Beijing 100044, China (e-mail: boai@bjtu.edu.cn).}
    \thanks{Octavia A. Dobre is with the Faculty of Engineering and Applied Science, Memorial University, St. John’s, NL A1C 5S7, Canada (E-mail: odobre@mun.ca).}
    \thanks{Arumugam Nallanathan is with the School of Electronic Engineering and Computer Science, Queen Mary University of London, London and also with the Department of Electronic Engineering, Kyung Hee University, Yongin-si, Gyeonggi-do 17104, South Korea (e-mail: a.nallanathan@qmul.ac.uk).}
    \thanks{{The code is available at https://github.com/XinkeXie-bjtu/PAsecure.}}
    }
	
	\maketitle

	\begin{abstract}

    Pinching antennas (PAs) represent an emerging flexible-antenna paradigm, where movable PAs arranged along dielectric waveguides enable on-demand customization of wireless channel characteristics. This paper exploits PAs to enhance physical-layer security (PLS) by increasing the channel diversity gap between legitimate users (LUs) and eavesdroppers (Eves). We concentrate on a dual-waveguide PA system, in which each waveguide is equipped with multiple PAs to serve multiple LUs under wiretapping from an Eve. Two  waveguide placement schemes are studied: parallel placement and orthogonal placement. To capture practical waveguide propagation effects, we adopt two in-waveguide channel models: in-waveguide phase shifts, and in-waveguide phase shifts and attenuation. We formulate secure sum rate (SSR) and secrecy energy efficiency (SEE) maximization problems, and develop a two-stage optimization framework to tackle these non-convex problems. In the first stage, we design a feasibility-enhanced particle swarm optimization (PSO) algorithm named FeaPSO to search for valid PA positions complying with minimum inter-PA spacing constraints. The second stage leverages successive convex approximation (SCA) to jointly optimize baseband beamforming and artificial noise vectors. Simulation results validate the effectiveness of the proposed two-stage algorithm and confirm that PA adjustment yields substantial gains in both SSR and SEE. Moreover, the simulation outcomes  demonstrate the performance benefits brought by orthogonal waveguide placement for PLS enhancement.

	\end{abstract}

	
	\IEEEpeerreviewmaketitle

	\setlength{\parindent}{1em}
	\section{Introduction}

\subsection{Background}

Physical-layer security (PLS) has emerged as an effective paradigm for information security, which leverages the channel diversity between legitimate users (LUs) and eavesdroppers (Eves) to realize secure transmission \cite{bckpls1}. However, PLS performance degrades under limited spatial degrees of freedom (DoF) at transceivers or when Eves experience favorable channel conditions \cite{bckpls2}. For instance, a base station (BS) struggles to deliver high-quality secure coverage for far-field LUs if Eves are located in the vicinity of the BS. More critically, PLS may become infeasible when Eves lie on the line-of-sight path between the BS and LUs. In conventional wireless systems, transceivers cannot actively manipulate channel characteristics, making such challenges intractable. Recently, emerging flexible-antenna technologies, including reconfigurable intelligent surfaces (RIS) \cite{ris}, movable-antenna systems \cite{zhu_ma} and fluid-antenna systems (FAS) \cite{wong_fas}, have enabled programmable channel state information (CSI). This capability enables the tailoring of partial CSI to enhance wireless coverage. As a representative example, RIS constructs alternative propagation paths between the BS and LUs to bypass Eves and mitigate interception, albeit introducing multiplicative fading. Meanwhile, path loss dominates the propagation of radio-frequency (RF) signals. Cell-free networks \cite{CF} represent another promising architecture that substantially boosts spectral efficiency by deploying access points (APs) in proximity to users. Even so, cell-free networks cannot reconfigure wireless CSI after deployment, which limits their ability to suppress strong wiretap channels of Eves. Consequently, existing flexible-antenna technologies and cell-free networks designed for PLS enhancement either incur extra path loss or struggle to address challenging wiretap scenarios.

Fortunately, the pinching antenna (PA) system (known as PASS) has been proposed \cite{Suzuki_pa,ding_FA}, which is able to provide large-scale and on-demand adjustments to wireless CSI \cite{pasurvey,pinch2O}, thereby offering an extra DoF for enhancing PLS. Specifically, PAs are realized  by placing  small dielectric particles, e.g., plastic pinches, onto a dielectric waveguide \cite{zhang_NF}. The waveguides can be deployed over a large scale, allowing PAs to be moved close to LUs to mitigate path loss as well as construct new links bypassing Eves, even when Eves are positioned in front of LUs. Thus, PAs exhibit significant potential to jointly enhance the channel conditions for LUs and PLS across most application scenarios. The key advantages of PAs have been demonstrated in most recent studies. For example, in \cite{PAISAC}, PAs were employed to enhance  integrated sensing and communications (ISAC), which significantly  mitigates the  communication-sensing trade-off. The authors in \cite{R3-2} further demonstrated that rate-splitting multiple access integrated with PA systems yielded higher target estimation accuracy for ISAC. In \cite{PANOMA}, the authors studied PA-enabled non-orthogonal multiple access (NOMA) and demonstrated  that PAs outperform conventional fixed-position antenna systems in terms of the quality-of-service (QoS)-aware achievable rate. A consistent observation was drawn in \cite{R3-1}, which analyzed PA-enabled NOMA short-packet transmission under the finite blocklength regime. In \cite{PASWIPT}, PAs were validated to achieve notable performance gains for simultaneous wireless information and power
transfer (SWIPT) over multi-input multi-output (MIMO) systems by reconfiguring CSI to mitigate large-scale fading. In \cite{PAOTA}, the authors validated that the joint PA positioning and communication design can greatly enhance the over-the-air computation accuracy. Existing studies validate the effectiveness of PAs  in customizing CSI to improve wireless communication services. 

Nevertheless, the movement range of PAs is inherently bounded by the fixed-placement waveguides. Specifically, when all waveguides are deployed in parallel, every PA is restricted to one single linear direction of movement. To circumvent this fundamental restriction, a straightforward solution is to explore alternative waveguide placement schemes. Beyond parallel waveguide placements, orthogonal waveguide placements introduce an additional orthogonal spatial dimension for PA adjustment. This extra spatial DoF is highly valuable for PLS, as PLS performance fundamentally relies on amplifying the statistical channel disparity between LUs and Eves.


\begin{table*}[t]
\centering
\footnotesize
\caption{Comparison with existing works on PAs for PLS.}\label{com}
\begin{tabular}{c||c|c|c|c|c|c}
\hline
Ref. & Waveguide Placement & PAs per Waveguide & LU & In-waveguide Channel Model & AN & Objective Function    \\
\hline\hline
\cite{Wang_papls} & Single & Multiple & Single &  In-waveguide phase shifts & $\times$ & SR \\
\hline
\cite{P_papls} & Multiple, parallel & Single & Single & In-waveguide phase shifts & $\checkmark$ & SR \\
\hline
\cite{Zhu_papls} &  Multiple, parallel & Multiple & Single & In-waveguide phase shifts & $\checkmark$ & SR \\
\hline
\cite{Illi_papls} & Multiple, parallel & Multiple & Single & In-waveguide phase shifts & $\checkmark$ & \thead{SR-constrained\\ illumination power} \\
\hline
\cite{sun_papls} & Multiple, parallel & Multiple & Multiple & In-waveguide phase shifts & $\times$& SSR \\
\hline
\cite{shan_pass} & Multiple, parallel & Multiple & Multiple & In-waveguide phase shifts & $\times$& Max-min SSR \\
\hline
Our work & 
{Dual, parallel, orthogonal} & Multiple&Multiple&\thead{In-waveguide phase shifts,\\In-waveguide phase shifts and attenuation} & $\checkmark$ & SSR, SEE\\
\hline
\end{tabular}
\end{table*}

\subsection{Contributions}

This paper investigates the PA-enabled PLS. Specifically, we consider a dual-waveguide scenario, where each waveguide  is equipped with multiple PAs to serve multiple LUs in the presence of an Eve.   The main contributions of this work are summarized as follows.

\begin{itemize}
    \item We formulate secure sum rate (SSR) and secure energy efficiency (SEE) maximization problems for the dual-waveguide PA system via joint optimization over PA positions, baseband beamforming, and artificial noise (AN) vectors.
Distinct from existing PA-enabled PLS studies that employ the parallel waveguide placement scheme, this paper also investigates the orthogonal waveguide placement scheme. In addition, two in-waveguide channel models are considered \cite{xu_iw}, i.e., in-waveguide phase shifts and in-waveguide phase shifts and attenuation.
    
    \item We propose a customized two-stage algorithm to tackle the formulated non-convex optimization problems. The first stage leverages particle swarm optimization (PSO) to search {high-quality feasible} PA positions, while the second stage adopts successive convex approximation (SCA) to jointly optimize beamforming and AN vectors for SSR and SEE maximization in a unified manner. 
    Specifically, we design a dedicated feasibility adjustment module embedded within PSO to always generate feasible PA configurations satisfying  minimum inter-PA spacing constraints. We also develop refined first-order Gaussian approximations for SCA iterations, which eliminate the need for semidefinite relaxation (SDR) and the Dinkelbach method required for fractional objectives. Furthermore, we numerically compare two waveguide  placement schemes based on average squared LU-waveguide distances.
    
    \item We carry out extensive numerical simulations to verify the performance superiority of PA systems over traditional fixed-position antenna systems, from which we draw four key insights. First, we validate the effectiveness of the proposed algorithm in comparison to the extensive search method. Second, we compare the two placement schemes, and the results show the necessity of the orthogonal waveguide placement. Third, we quantify and analyze the impacts of power budget on PLS performance. Fourth, we illustrate the influence of the in-waveguide attenuation on channel model. 
\end{itemize}




 The remainder of this paper is structured as follows. Section II provides a brief survey of related works. Section III  introduces the dual-waveguide-multi-PA system model and formulates SSR and SEE maximization problems. Section IV proposes a two-stage algorithm based on PSO and SCA methods, and provides a numerical comparison of two waveguide placement schemes.  Section V contains the simulations and discussions of the proposed algorithm to evaluate the dual-waveguide PA-enabled PLS. Section VI concludes this paper.

{\it Notations:} In this paper, $x$, ${\bf x}$, ${\bf X}$ and ${\cal X}$ respectively denote scalar, vector, matrix and set. ${\rm Re}\{ \cdot  \}$ is the real part of a complex-valued scalar, vector or matrix. $|| \cdot ||$ denotes the Euclidean norm for a complex vector and $| \cdot |$ is the magnitude for a complex scalar. $(\cdot)^T$ and $(\cdot)^H$ represent the transpose and conjugate transpose, respectively. ${\mathbb C}^M$ and  ${\mathbb C}^{M \times N}$ denote the set of $M \times 1$ complex-valued vectors and $M \times N$ complex-valued matrices, respectively. ${\bf I}_N$ and ${\bf 0}_N$ are the $N$-dimension identity matrix and zero vector, respectively. ${\bf a} \sim \mathcal{CN}({\bm \mu}, {\bm \Sigma})$ denotes that ${\bf a}$ is a complex-valued circularly symmetric Gaussian random variable with mean  ${\bm \mu}$ and covariance matrix ${\bm \Sigma}$. ${a} \sim \mathcal{U}(a_{1},a_{2})$ denotes that ${a}$ follows a uniform distribution over the interval $[a_{1},a_{2}]$

\section{Related Work}

To date, there have been a few related studies on PA systems. Below, we briefly summarize related works on  optimization techniques for PAs and PA-enabled PLS.

\subsection{Optimization Techniques for PAs}

Optimizing PA systems poses significant challenges, primarily due to two key factors: the strong coupling between PA positions and transmission parameters, and the presence of exponential functions associated with PA positions \cite{pinch1O}. For most single-waveguide-single-user cases,  closed-form optimal or near-optimal solutions can be derived. In \cite{PAISAC}, a closed-form solution was derived for sensing requirement constrained rate maximization in the scenario of a single transmitting waveguide serving one user and one target. The rate maximization problem for downlink single-waveguide-single-user systems was relaxed into two subproblems with closed-form solutions, where one minimizes large-scale path loss, and another  maximizes received signal strength. 

However, the technical challenges of optimizing PA systems become even more pronounced in scenarios with multiple waveguides and  multiple users. Existing optimization techniques can be categorized into three categories: alternating optimization (AO)-based convex optimization\footnote{The  block coordinate
 descent (BCD)-based algorithm also falls under this category.} (CVXopt), search-aided CVXopt, and learning-based optimization. Particularly, the AO-based or search-aided CVXopt is primarily designed for exploring system insights, while learning-based optimization is focused on fast real-time implementation. For example, in \cite{Bereyhi_pasum}, the authors studied the problem of multi-user detection and beamforming design in the uplink and downlink of  PA-MIMO systems, with the aim of maximizing the weighted sum rate. The problem was solved using  a BCD-based approach, and the results indicated that the throughput in both uplink and downlink was boosted by PAs,  compared to conventional MIMO architectures. In \cite{ding_FA}, the authors employed a search-based algorithm combined  with zero-forcing (ZF) beamformers to derive an upper bound on max-min signal-to-interference-plus-noise ratio (SINR) of two users served by a multi-waveguide-multi-PA system.  In \cite{GPASS}, the authors proposed a graph neural network (GNN) enabled joint PA positioning and beamforming design to maximize the sum rate for multi-waveguide-multi-user systems, which demonstrates high spectral efficiency alongside low inference complexity. Similarly, in \cite{xie_gnn}, the authors proposed a bipartite graph attention network (GAT)-based algorithm to  maximize  energy efficiency for downlink PA systems, which scales with the number of users.

\subsection{PA-Enabled PLS}

In \cite{B_papls}, the authors considered a single-waveguide-single-PA system to transmit confidential information to a single-antenna LU in the presence of an Eve. They analyzed the system performance in terms of average secrecy capacity, strictly positive secrecy capacity, and secrecy outage probability, and derived mathematical expressions for these metrics. In \cite{Wang_papls}, the authors modeled a single-waveguide-multi-PA system within a rectangular region. They considered  PLS-oriented transmission for a transceiver in the presence of an Eve, and formulated a secrecy rate maximization problem. This problem was then re-formulated as a coalitional game with non-transferable utility, aiming to identify the optimal combination of activated PAs. The individual impact of each PA was quantified based on the Shapley value and marginal contribution, thereby providing a fair and efficient method for performance evaluation.

In addition to single-waveguide scenarios, multi-waveguide scenarios offer more design flexibility while introducing inherent challenges. In \cite{P_papls}, the authors employ PAs to enhance PLS over fixed-position antennas. They consider the multi-waveguide-single-antenna system, and formulate an AN{\footnote{As proven for fixed-position antenna systems \cite{an}, the AN-aided design delivers SSR or SEE performance no inferior to that of the AN-free design. This conclusion can be extended to PA systems. If PAs are flexibly configured to make LUs' CSI vectors fully orthogonal to all Eves' CSI vectors, AN yields no additional performance improvement.}}-aided secrecy rate maximization problem for an LU and an Eve. To address the considered problem, an AO-based algorithm is proposed with transmit beamformers (i.e., beamforming vectors and AN matrix) and PA positions being alternatively optimized until convergence. Particularly, they optimized the transmit beamformers with a slack variable and an AO framework; and they utilized an exhaustive grid-search strategy to configure the PA positions. In \cite{Zhu_papls}, the authors investigated AN-aided PLS  in multi-waveguide scenarios, aiming to maximize the secrecy rate for one LU. Two transmission architectures were proposed: waveguide division (WD) and waveguide multiplexing (WM), distinguished by whether each waveguide carries only one type of signal or mixed signals. The WD case was solved using the projection approximation subspace tracking algorithm for pinching beamforming and an SCA method for baseband power allocation via an AO method. The WM case was addressed by designing pinching beamforming via PSO and baseband beamforming via the SCA method.  In \cite{Illi_papls}, a PA-ISAC system was studied,  where a dual-functional PA system serves multiple LUs while sensing a set of malicious targets. They first determined the optimal PA positions, and then, optimized the legitimate signal beamforming and AN covariance matrices, with the aim of  maximizing sensing performance, while satisfying information security requirement. In \cite{sun_papls}, the authors considered a multi-waveguide-multi-PA system, where PAs transmit information signals to multiple LUs in the presence of several Eves. They formulated a weighted SSR maximization problem subject to power budget  and PA positioning constraints, and the Lagrange dual transform, fractional programming, BCD method are jointly employed to solve it. In \cite{shan_pass}, the authors modeled a downlink secure multicast transmission system powered by PAs, where each multicast group requires the same information signals. They formulated a max-min SSR problem and solved it under a majorization-minimization framework. 

The effectiveness of PA in enhancing PLS has been extensively validated in the aforementioned works. However, all existing literature is limited to single-axis parallel waveguide placement, where PAs are only adjustable along one shared coordinate dimension.  Moreover, existing works also only adopt the simplified in-waveguide phase-shift-only channel model and merely target SSR maximization. For clarity, we highlight the core differences between these existing works and our work in Table \ref{com}. 




\section{System Model}

   \begin{figure} [t]
	\centering
	\subfloat[\label{1a}]{		\includegraphics[ width=.49\textwidth]{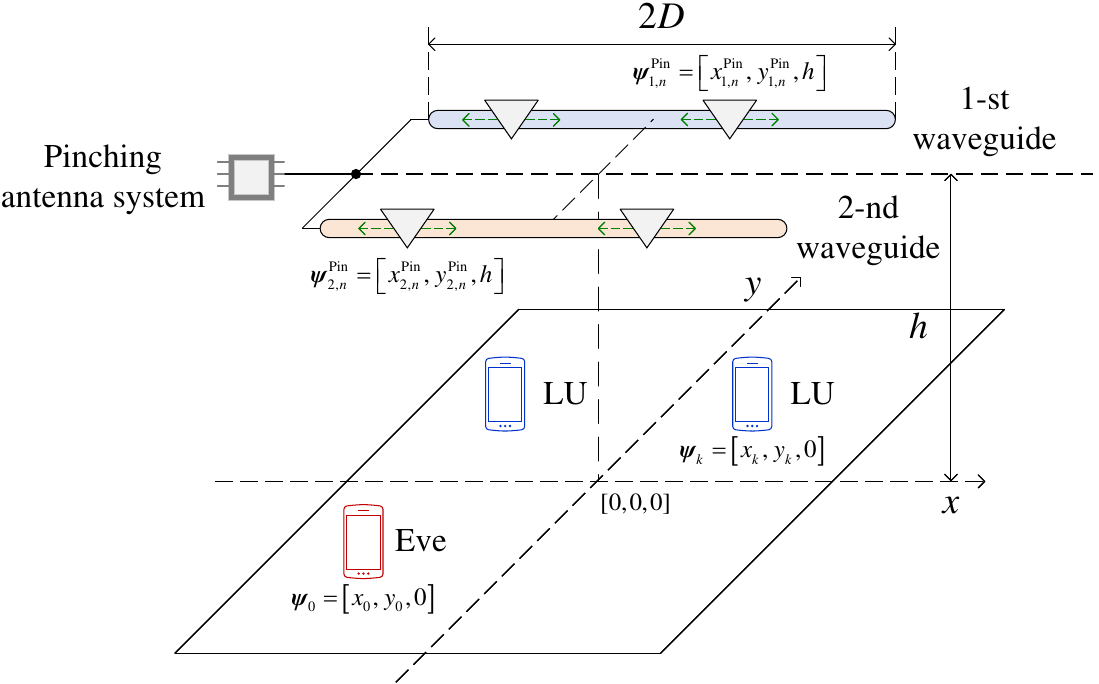}}
	\\
	\subfloat[\label{1b}]{		\includegraphics[ width=.375\textwidth]{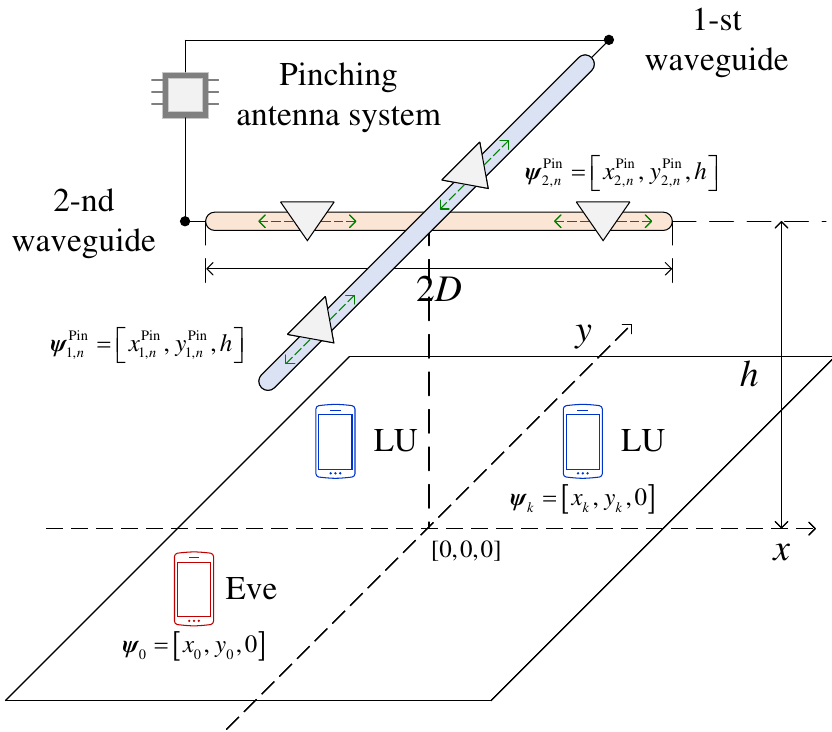} }
	\caption{Illustration of dual-waveguide PA system: (a) parallel placement; (b) orthogonal placement. The two waveguides deploy multiple PAs, and they are placed parallelly or orthogonally to serve multiple LUs in the presence of an Eve.  }
	\label{sys} 
\end{figure}

\begin{table}[t]\label{table:pare}
	\caption{Summary of Main Notations Used in System Model}
	\centering
	\begin{tabular}{c|c}
		\hline
		{\bf Notation} & {\bf Definition} \\
		\hline\hline
	      $K$ & Number of LUs. \\
        \hline
        $N$ & Number of PAs per waveguide \\
        \hline
            $2D$ &  {Side length of the square serving area}   \\
        \hline
        $h$ & Height of waveguides\\
        \hline
           \multirow{2}{*}{${\bm \psi}_{k}\in{\mathbb R}^3$}  &   Position of the $k$-th LU ($k\ne0$) \\
           \cline{2-2}
             &   Position of Eve ($k=0$) \\
        \hline
        ${\bm\psi}^{\rm Pin}_{l,n}\in{\mathbb R}^3$ & Position of the $n$-th PA on the $l$-th waveguide \\
                \hline
        ${\bm\psi}^{\rm Pin}_{l,0}\in{\mathbb R}^3$ &  Position of feed point of the
$l$-th waveguide\\
        \hline
           \multirow{2}{*}{$s_k\in{\mathbb C}$,${\bf w}_k\in{\mathbb C}^2$}  &   \thead{Information signal, baseband beamforming\\vector for the $k$-th LU ($k\ne0$)} \\
           \cline{2-2}
             &  AN signal, baseband AN vector ($k=0$) \\
        \hline
        ${\bf G}(\{{\bm \psi}^{\rm Pin}_{l,n}\})\in{\mathbb C}^{2N\times 2}$ & Pinching beamforming matrix \\
        \hline
        ${\bf g}(\{{\bm \psi}^{\rm Pin}_{l,n}\})\in{\mathbb C}^N$ & In-waveguide propagation \\
        \hline
        $\lambda_{\rm G}$ & Guided wavelength\\
        \hline
        $\zeta$ & In-waveguide attenuation coefficient\\
        \hline
        ${\bf h}_{k}(\{{\bm \psi}^{\rm Pin}_{l,n}\})\in{\mathbb C}^{2N}$ & CSI vector from PAs to the $k$-th LU/Eve\\
        \hline
        $y_k$, $\omega_k$ &Received signal, AWGN at the $k$-th LU/Eve \\
        \hline
        ${\bf f}_k(\{{\bm\psi}_{l,n}^{\rm Pin}\})\in{\mathbb C}^2$  & Representation of ${\bf h}^T_{k}(\{{\bm\psi}_{l,n}^{\rm Pin}\}){\bf G}(\{{\bm\psi}_{l,n}^{\rm Pin}\})$\\
        \hline
	\end{tabular}
\end{table}

  As shown in Figure \ref{sys}, we consider a PA system that serves $K$ LUs in the presence of an Eve. The serving area is a square of size $2D$, defined in a 3D coordinate system.  Denote the positions of LUs/Eve as $\{{\bm \psi}_{k}=[x_k,y_k,0]\}_k$ which are fixed. Here, $k\in\{1,2,\ldots,K\}$ and $k=0$ represent the $k$-th LU and Eve, respectively. To enable secure information transmission, the system generates the information and the AN signals. Particularly, we focus on a special case of multi-waveguide PA systems, namely dual-waveguide PA systems. There are $N$ PAs that can be flexibly moved along the waveguide over a large scale of $[-D,D]$, where $2D$ represents  the length of each waveguide. Denote the position of the $n$-th ($n\in{\cal N}\triangleq[1,2,\ldots,N]$) PA on the $l$-th waveguide as ${\bm\psi}^{\rm Pin}_{l,n}=[{x}^{\rm Pin}_{l,n},y^{\rm Pin}_{l,n},h]$ where $l\in\{1,2\}$ represents index of the two waveguide and $h$ represents the height of the waveguides. Two waveguide placement schemes are considered for the dual-waveguide PA system, namely parallel placement (Figure \ref{sys}(a)) and orthogonal placement (Figure \ref{sys}(b)): 
  \begin{itemize}
      \item {\emph{Parallel placement}.} Two waveguides are deployed in parallel at a distance of $D/3$ between them. Besides, $y^{\rm Pin}_{l,n}$ and $h$ are pre-given with  ${x}^{\rm Pin}_{l,n}\in[-D,D]$.
      \item {\emph{Orthogonal placement}.} Two waveguides are deployed orthogonally, with their intersection point being the origin. Besides, $x^{\rm Pin}_{1,n}$, $y^{\rm Pin}_{2,n}$ and $h$ are pre-given with $y^{\rm Pin}_{1,n},x^{\rm Pin}_{2,n}\in[-D,D]$.
  \end{itemize}

The constructed system adopts ideal hardware assumptions, in which PA adjustment latency and control signaling overhead are neglected, and CSI is assumed stationary throughout the channel coherence time. To accommodate practical communication scenarios, the BS can extend the update cycle of movable PAs and retain PA configurations across multiple coherence slots to reduce adjustment latency and signaling overhead, as exemplified by two-timescale frameworks \cite{two-timescale}.  {We further note that these assumptions yield theoretical throughput that exceeds the effective practical throughput.}

  \subsection{Baseband Beamforming and Pinching Beamforming}

Denote $\{s_k\}$ as the generated signals, where $s_k$ ($k>0$) is for the $k$-th user and $s_0$ is the AN signal. Without loss of generality, we assume that ${\mathbb E}\{|s_k|^2\}=1$. The signals after baseband beamforming and pinching beamforming (i.e., leaving PAs) are given by 
\begin{flalign}
    {\bf s} = \sum\nolimits_{k=0}^K{\bf G}\left(\left\{{\bm \psi}^{\rm Pin}_{l,n}\right\}\right){\bf w}_{k}s_{k}\in{\mathbb C}^{2N},
\end{flalign}
where ${\bf G}(\{{\bm \psi}^{\rm Pin}_{l,n}\})\in{\mathbb C}^{2N\times 2}$ denotes the pinching beamforming matrix, and ${\bf w}_{k}=[w_{k1},w_{k2}]^T\in{\mathbb C}^{2}$  denotes the baseband beamforming vector ($k>0$) or AN vector ($k=0$). Here, ${\bf G}(\{{\bm \psi}^{\rm Pin}_{l,n}\})$ can be expressed as 
\begin{flalign}\label{pinchingbeam}
    {{\bf G}\left(\left\{{\bm \psi}^{\rm Pin}_{l,n}\right\}\right)=\left[\begin{array}{cc}
         {\bf g}\left(\left\{{\bm \psi}^{\rm Pin}_{1,n}\right\}\right) &{\bf 0}_{N}  \\
         {\bf 0}_N & {\bf g}\left(\left\{{\bm \psi}^{\rm Pin}_{2,n}\right\}\right)
    \end{array}\right],}
\end{flalign}
where 
\begin{flalign}
    {\bf g}&\left(\{{\bm \psi}^{\rm Pin}_{l,n}\}\right)=\\
    &\sqrt{\frac{1}{N}}\left[{g}\left({\bm \psi}^{\rm Pin}_{l,1}\right),{g}\left({\bm \psi}^{\rm Pin}_{l,2}\right),\ldots,{g}\left({\bm \psi}^{\rm Pin}_{l,N}\right)\right]^T\in{\mathbb C}^N,\nonumber
\end{flalign} represents the in-waveguide propagation response. Here, the term \(\sqrt{1/N}\) indicates that each PA radiates an equal proportion of the power allocated to the corresponding waveguide \cite{11263923}. Specifically, depending on whether the in-waveguide attenuation is neglected or not, we consider two models of ${\bf g}(\{{\bm \psi}^{\rm Pin}_{l,n}\})$:
\begin{itemize}
    \item \emph{In-waveguide phase shifts}. This case omits the propagation loss within the waveguide, as some references indicate the in-waveguide attenuation is negligible under some conditions \cite{xu_iw,xu_iw2}. Then,  ${\bf g}(\{{\bm \psi}^{\rm Pin}_{l,n}\})$ only reflects the phase shifts of  in-waveguide signal propagation, and we have  
    \begin{flalign}
        g\left({\bm\psi}^{\rm Pin}_{l,n}\right)=\exp\left(-j\frac{2\pi}{\lambda_{\rm G}}\left\|{\bm\psi}^{\rm Pin}_{l,0}-{\bm\psi}^{\rm Pin}_{l,n}\right\|\right),
    \end{flalign}
    where ${\bm\psi}^{\rm Pin}_{l,0}$ denotes the position of feed point of the $l$-th waveguide, and $\lambda_{\rm G}=\lambda/n_{\rm neff}$ denotes the guided wavelength with $\lambda$ being the free-space wavelength  of the carrier frequency and $n_{\rm neff}$ being the effective refractive index of a dielectric waveguide.
    \item \emph{In-waveguide phase shifts and attenuation}. This case models both power loss and phase shifts of signals propagated within the waveguide. The in-waveguide signal propagation power loss of the $n$-th PA in the $l$-th waveguide is represented by\cite{ref_wave}
    \begin{flalign}\label{ac}
        \exp\left(-\zeta\left\|{\bm\psi}^{\rm Pin}_{l,0}-{\bm\psi}^{\rm Pin}_{l,n}\right\|\right),
    \end{flalign}
    where $\zeta$ denotes the attenuation coefficient.
    Then, we have  
    \begin{flalign}
        &g\left({\bm\psi}^{\rm Pin}_{l,n}\right)=\\
        &\exp\left(-\zeta\left\|{\bm\psi}^{\rm Pin}_{l,0}-{\bm\psi}^{\rm Pin}_{l,n}\right\|-j\frac{2\pi}{\lambda_{\rm G}}\left\|{\bm\psi}^{\rm Pin}_{l,0}-{\bm\psi}^{\rm Pin}_{l,n}\right\|\right).\nonumber
    \end{flalign}
\end{itemize}

  \subsection{Channel Model and Information Rate}

The CSI vector{\footnote{{We assume perfect CSI of Eve, which serves as a conservative benchmark for secrecy performance and yields an upper bound on the SSR and SEE for cases with imperfect Eve CSI, including both bounded CSI errors and statistical CSI errors.
}}} from all PAs to the $k$-th LU/Eve is given by
\begin{flalign}\label{channel}
	 {\bf h}^H_{k}&\left(\left\{{\bm \psi}^{\rm Pin}_{l,n}\right\}\right) \triangleq  
     \left[ \frac{{\sqrt \eta  {\exp\left({ { - j\frac{2\pi}{{\lambda }} \left\| {{{\bm\psi} _k} - {\bm\psi} _{1,1}^{{\rm{Pin}}}} \right\|} }\right)}}}{{\left\| {{{\bm\psi} _k} - {\bm\psi}_{1,1}^{{\rm{Pin}}}} \right\|}}, \ldots, \right.\nonumber\\
     &\left.\frac{{\sqrt \eta  {\exp\left({ { - j\frac{2\pi}{{\lambda }} \left\| {{{\bm\psi} _k} - {\bm\psi} _{2,N}^{{\rm{Pin}}}} \right\|} }\right)}}}{{\left\| {{{\bm\psi} _k} - {\bm\psi}_{2,N}^{{\rm{Pin}}}} \right\|}} \right]^T\in{\mathbb C}^{1\times2N},
\end{flalign}
where $\eta=\lambda^2/(4\pi)^2$.  

The received signal at the $k$-th LU/Eve is given by 
\begin{flalign}
&y_{k} =\sum\nolimits_{k^{\prime}=0}^K{\bf h}^H_{k}\left(\left\{{\bm\psi}_{l,n}^{\rm Pin}\right\}\right){\bf G}\left(\left\{{\bm\psi}_{l,n}^{\rm Pin}\right\}\right){\bf w}_{k^{\prime}}s_{k^{\prime}} + \omega_k,
\end{flalign}
where $\omega_k$ denotes the additive white Gaussian noise (AWGN) with power of $\sigma_k^2$. Here, for notational simplicity, we denote 
\begin{flalign}\label{f_psi}
    {\bf f}_k\left(\{{\bm\psi}_{l,n}^{\rm Pin}\}\right)\triangleq {\bf h}^H_{k}\left(\left\{{\bm\psi}_{l,n}^{\rm Pin}\right\}\right){\bf G}\left(\left\{{\bm\psi}_{l,n}^{\rm Pin}\right\}\right)\in{\mathbb C}^{1\times 2},
\end{flalign}
such that $y_{k}$ is represented as 
\begin{flalign}
    y_{k} = \sum\nolimits_{k^{\prime}=0}^K{\bf f}_{k}\left(\left\{{\bm\psi}_{l,n}^{\rm Pin}\right\}\right){\bf w}_{k^{\prime}}s_{k^{\prime}} + \omega_k.
\end{flalign}

Then, the information rate for receiving $s_k$  at the $k$-th LU is given by
\begin{flalign}
    &R_k\left(\left\{{\bm\psi}_{l,n}^{\rm Pin},{\bf w}_k\right\}\right)=\nonumber\\
    &\log_2\left(1+\frac{\left|{\bf f}_k\left(\{{\bm\psi}_{l,n}^{\rm Pin}\}\right){\bf w}_k\right|^2}{\sum\nolimits_{k^{\prime}=0,k^{\prime}\ne k}^K\left|{\bf f}_k\left(\{{\bm\psi}_{l,n}^{\rm Pin}\}\right){\bf w}_{k^\prime}\right|^2+\sigma^2_k} \right),
\end{flalign}
and the leakage to Eve is given by
\begin{flalign}
    &R^{\rm Eve}_k\left(\left\{{\bm\psi}_{l,n}^{\rm Pin},{\bf w}_k\right\}\right)=\nonumber\\
    &\log_2\left(1+\frac{\left|{\bf f}_0\left(\{{\bm\psi}_{l,n}^{\rm Pin}\}\right){\bf w}_k\right|^2}{\sum\nolimits_{k^{\prime}=0,k^{\prime}\ne k}^K\left|{\bf f}_0\left(\{{\bm\psi}_{l,n}^{\rm Pin}\}\right){\bf w}_{k^\prime}\right|^2+\sigma^2_0} \right).
\end{flalign}
The corresponding secure information rate is defined by
\begin{flalign}
&R^{\rm Sec}_k\left(\left\{{\bm \psi}^{\rm Pin}_{l,n},{{\bf w}_{k}}  \right\}\right) = \\
&\left[ R_k\left(\left\{{\bm\psi}_{l,n}^{\rm Pin},{\bf w}_k\right\}\right) - R^{\rm Eve}_k\left(\left\{{\bm\psi}_{l,n}^{\rm Pin},{\bf w}_k\right\}\right)\right]^+.\nonumber
\end{flalign}

  \begin{figure}[t]
    \centering
    \includegraphics[width=0.48\textwidth]{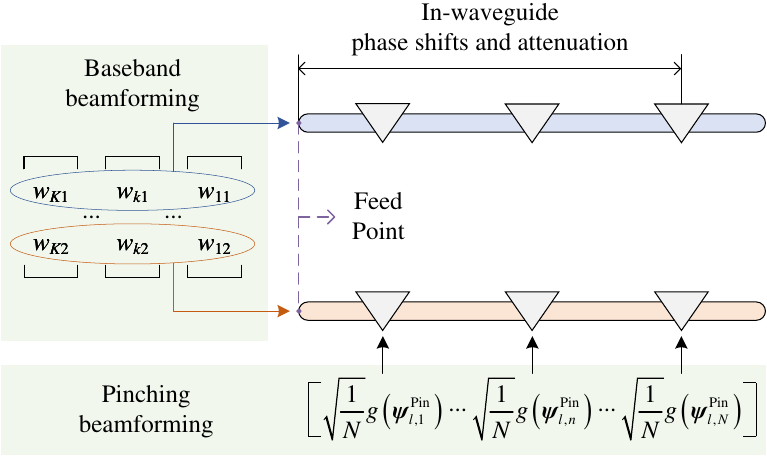}
    \caption{Illustration of baseband beamforming and pinching beamforming.} 
    \label{fig:PA}
\end{figure}


\subsection{Problem Formulation}

Our goal is to maximize SSR or SEE for the considered system, which is formulated by 
\begin{subequations}
	\begin{flalign}
		&{{\bf P}_1}: \max_{\left\{{\bm \psi}^{\rm Pin}_{l,n},{{\bf w}_{k}}  \right\}}\sum\nolimits_{k=1}^K R_{\rm Sec}\left(\left\{{\bm \psi}^{\rm Pin}_{l,n},~{{\bf w}_{k}}  \right\}\right)\label{p1}\\
       & {{\bf P}_2}:  \max_{\left\{{\bm \psi}^{\rm Pin}_{l,n},{{\bf w}_{k}}  \right\}} \frac{\sum\nolimits_{k=1}^K R_{\rm Sec}\left(\left\{{\bm \psi}^{\rm Pin}_{l,n},~{{\bf w}_{k}}  \right\}\right)}{\sum\nolimits_{k^{\prime}=0}^K\left\|{\bf w}_{k^{\prime}}\right\|^2 + P_{\rm C}}\label{p2}\\
        {\rm s.t.}~& \sum\nolimits_{k=0}^K\left\|{\bf w}_k\right\|^2 \le P_{\max},\label{cons:1}\\
        &\left\{ \begin{array}{l}
x_{l,n}^{{\rm{Pin}}} \in [ - D,D],~l \in \left\{ 1,2 \right\},~n \in {\cal N}\\
x_{l,n}^{{\rm{Pin}}} - x_{l,n - 1}^{{\rm{Pin}}} \ge \Delta ,l \in \left\{ 1,2 \right\},n \in {\cal N}/\left\{ 1 \right\}
\end{array} \right.,\label{cons:pd}\\
{\rm or}~&\left\{ \begin{array}{l}
y_{1,n}^{{\rm{Pin}}},x_{2,n}^{{\rm{Pin}}} \in [ - D,D],~n \in {\cal N}\\
y_{1,n}^{{\rm{Pin}}} - y_{1,n-1}^{{\rm{Pin}}} \ge \Delta,~n \in {\cal N}/\left\{ 1 \right\}\\
x_{2,n}^{{\rm{Pin}}} - x_{2,n-1}^{{\rm{Pin}}} \ge \Delta,~n \in {\cal N}/\left\{ 1 \right\}
\end{array} \right.,\label{cons:od}
	\end{flalign}
\end{subequations}
where $P_{\max}$ and $P_{\rm C}$ denote the power budget and circuit power  of the BS, respectively, and $\Delta$ is the minimum guard distance to avoid the antenna coupling. Constraints \eqref{cons:pd} and \eqref{cons:od} are due to  parallel placement and  orthogonal placement, respectively. Note that \eqref{cons:pd} and \eqref{cons:od} imply that the pinching antennas on one waveguide are deployed in an ascending order.

\section{Proposed Two-Stage Algorithm}

The considered problems are non-convex and hard to solve due to the deeply coupled variables. Particularly, there are exponential terms in objective functions (introduced by pinching beamforming matrix and CSI vector). To solve the considered problem, we propose a two-stage algorithm, which first determines the PA positions and then optimizes the baseband beamforming vectors. 


\subsection{First Stage: Pinching-Antenna Placement}

We design the first stage to find a {high-quality feasible} PA position with given $\{\widetilde{\bf w}_k\}$. However, since the information rates involve sine functions of 
 $\{{\bm\psi}^{\rm Pin}_{l,n}\}$, gradient-based optimization methods may be inapplicable. Subsequently, we adopt a PSO-based algorithm to optimize $\{{\bm\psi}^{\rm Pin}_{l,n}\}$. Specifically, we integrate the PSO-based algorithm with a feasibility adjustment module to ensure the generation of feasible solutions, which is termed FeaPSO. Notably, as $\{\widetilde{\bf w}_k\}$ are strongly dependent on $\{{\bm\psi}^{\rm Pin}_{l,n}\}$, we need to fine-tune $\{\widetilde{\bf w}_k\}$ accordingly based on the updated $\{{\bm\psi}^{\rm Pin}_{l,n}\}$.

Without optimizing $\{\widetilde{\bf w}_k\}$, the two problems under consideration are equivalent to 
\begin{subequations}
	\begin{flalign}
		&{{\bf P}_{\rm 1A^{\prime}}}: \max_{\left\{{\bm\psi}^{\rm Pin}_{l,n}\right\}}\sum\nolimits_{k=1}^K R_{\rm Sec}\left(\left\{{\bm \psi}^{\rm Pin}_{l,n},~{\widetilde{\bf w}_{k}}  \right\}\right)\\
        {\rm s.t.}~&\eqref{cons:pd}~{\rm or}~\eqref{cons:od}.
	\end{flalign}
\end{subequations}

For notational simplicity, we take parallel placement (i.e., (\ref{cons:pd})) as an example. We define a particle swarm of size $I$ and initialize the $i$-th ($i\in\{1,2,...,I\}$) particle associated with the optimization variables as 
\begin{flalign}
    {\bf x}^{\rm Pin}{\left[i\right]} = \left[{x}^{\rm Pin}_{1,1}{\left[i\right]},{x}^{\rm Pin}_{1,2}{\left[i\right]},\ldots,{x}^{\rm Pin}_{2,N-1}{\left[i\right]},{x}^{\rm Pin}_{2,N}{\left[i\right]} \right]^T,
\end{flalign}
where ${x}^{\rm Pin}_{l,n}{[i]}$ denotes the $x$-coordinate of the $n$-th PA on the $l$-th waveguide in the $i$-th  particle.
 Correspondingly, we define the initial velocity of the $i$-th particle as
\begin{flalign}
    {\bf u}^{\rm Pin}{\left[i\right]} = \left[{u}^{\rm Pin}_{1,1}{\left[i\right]},{u}^{\rm Pin}_{1,2}{\left[i\right]},\ldots,{u}^{\rm Pin}_{2,N-1}{\left[i\right]},{u}^{\rm Pin}_{2,N}{\left[i\right]} \right]^T.
\end{flalign}
The PSO algorithm indicates that ${\bf u}^{\rm Pin}{\left[i\right]}$ can be updated by the following linear combination:
\begin{flalign}\label{update:u}
    {\bf u}^{\rm Pin}{\left[i\right]} \leftarrow  &\kappa{\bf u}^{\rm Pin}{\left[i\right]}+v_1\eta_1\left({\bf x}^{\rm P}{\left[i\right]}-{\bf x}^{\rm Pin}\left[i\right] \right)+\\
    &v_2\eta_2\left({\bf x}^{\rm G}-{\bf x}^{\rm Pin}\left[i\right] \right),\nonumber
\end{flalign}
where ${\bf x}^{\rm P}[i]$ and ${\bf x}^{\rm G}$ denote the personal best position of $i$-th particle and the global best position of the entire swarm, respectively, $\kappa$ denotes the inertia weight, $v_1$ and $v_2$ denote the acceleration coefficients, and $\eta_1, \eta_2 \sim \mathcal{U}(0,1)$ are random variables to introduce stochasticity. With ${\bf u}^{\rm Pin}{[i]}$, ${\bf x}^{\rm Pin}[i]$ is updated by 
\begin{flalign}\label{update:x}
    {\bf x}^{\rm Pin}\left[i\right]\leftarrow  & {\bf x}^{\rm Pin}\left[i\right]+{\bf u}^{\rm Pin}{\left[i\right]}.
\end{flalign}

To satisfy (\ref{cons:pd}), we introduce the following \emph{feasibility adjustment module} \cite{xie_gnn}. Each element in ${\bf x}^{\rm Pin}[i]$ is first adjusted by
\begin{flalign}\label{ad:1}
    x^{\rm Pin}_{l,n}\left[i\right]=\left\{ \begin{array}{l}
D,~{\rm if}~x^{\rm Pin}_{l,n}\left[i\right]>D\\
-D,~{\rm if}~x^{\rm Pin}_{l,n}\left[i\right]<-D\\
x^{\rm Pin}_{l,n}\left[i\right],~{\rm otherwise}
\end{array} \right..
\end{flalign}
Then, sort the elements of $\{x^{\rm Pin}_{l,n}[i]\}_n$ in an ascending order such that ${\delta}_n\triangleq x^{\rm Pin}_{l,n}[i]-x^{\rm Pin}_{\tau,n-1}[i]\ge0$ $(\forall n\in{\mathcal N})$. We adjust ${\delta}_n$ by
\begin{flalign}\label{D_max}
   {{\delta}_{n} \leftarrow \frac{B_{\max}}{{\rm max}\left( B_{\max}, \sum\nolimits_{n^{\prime}=1}^{N}{\delta}_{n^{\prime}} \right)}{\delta}_n},
\end{flalign}
where $B_{\max}\triangleq 2D-( N-1 )\Delta$. With $\{\delta_n\}$, we obtain 
\begin{flalign}\label{ad:2}
x^{\rm Pin}_{l,n}[i]\leftarrow\left\{ \begin{array}{l}
{\delta}_n-D,~n=1\\
x^{\rm Pin}_{l,n}[i]+{\delta}_n+\Delta-D,~n\in{\cal N}\setminus\{1\} \end{array} \right..
\end{flalign}

With feasible ${\bf x}^{\rm Pin}{[i]}$, we obtain $\{{\bf f}_k(\{{\bm \psi}^{\rm Pin}_{l,n}\})\}_{\{x^{\rm Pin}_{l,n}= x^{\rm Pin}_{l,n}[i]\}}$ based on \eqref{f_psi}. We then set $\widetilde{\bf w}_{0}={\bf 0}_2$ and obtain $\{\widetilde{\bf w}_{k}\}_{k\ne 0}$ via the schemes\footnote{Note that  AN functions as interference to both LUs and Eve. Typically, the power of AN remains a very low level, as demonstrated in \cite{an}. As a result, $\widetilde{\bf w}_{0}$ is approximately set by ${\bf 0}_2$ in the first stage, and subsequently refined in the second stage.} in \cite{scheme} given by \eqref{w}, where $p_k = \lambda_k = {P_{\max}}/{K}$. Note that the obtained $\widetilde{\bf w}_{k}$ satisfy the power budget constraint \eqref{cons:1}. 

\begin{figure*}[b]
\hrule
\begin{flalign}\label{w}
\widetilde{\bf w}_k = \sqrt{p_k} \frac{\left({\bf I} +\sum\nolimits_{{{k^\prime}=1}}^K \frac{\lambda_{k^{\prime}}}{\sigma^2_k}{\bf f}^H_{{k^\prime}}\left(\{{\bm\psi}_{l,n}^{\rm Pin}\}\right){\bf f}_{{k^\prime}}\left(\{{\bm\psi}_{l,n}^{\rm Pin}\}\right) \right)^{-1}{\bf f}^H _{{k}}\left(\{{\bm\psi}_{l,n}^{\rm Pin}\}\right)}{\left\|\left({\bf I} +\sum\nolimits_{{{k^\prime}=1}}^K \frac{\lambda_{k^{\prime}}}{\sigma^2_k}{\bf f}^H_{{k^\prime}}\left(\{{\bm\psi}_{l,n}^{\rm Pin}\}\right){\bf f}_{{k^\prime}}\left(\{{\bm\psi}_{l,n}^{\rm Pin}\}\right)\right)^{-1}{\bf f}^H _{{k}}\left(\{{\bm\psi}_{l,n}^{\rm Pin}\}\right)\right\|}
\end{flalign}
\end{figure*}

The fitness function, which represents the achievable SSR, of each particle is given by
\begin{flalign}\label{fitness}
    &{\cal F}\left({\bf x}^{\rm Pin}\left[i\right]\right)=\sum\nolimits_{k=1}^K R_{\rm Sec}\left(\left\{{\bm \psi}^{\rm Pin}_{l,n},~{\widetilde{\bf w}_{k}}  \right\}\right)\left|_{\left\{x^{\rm Pin}_{l,n}= x^{\rm Pin}_{l,n}\left[i\right]\right\}}\right..
\end{flalign}

We summarize FeaPSO for solving Problem $\rm P_{1A^{\prime}}$ in Algorithm \ref{alg2}. Moreover, for clarity, we provide a flowchart to illustrate the FeaPSO workflow and highlight its key differences from traditional PSO, i.e., the mechanisms to guarantee feasible ${\{{\bm \psi}^{\rm Pin}_{l,n},{{\bf w}_{k}} \}}$.

\begin{algorithm}[!t]\label{alg2}
\caption{FeaPSO for solving Problem $\rm P_{1A^{\prime}}$.}
Initialize $\{{\bf x}^{\rm Pin}[i],{\bf u}^{\rm Pin}[i]]\}$ satisfying \eqref{cons:pd}\;
Evaluate the fitness function value with $\{{\bf x}^{\rm Pin}[i]\}$ to obtain the personal best positions $\{{\bf x}^{\rm P}[i]\}_i$ and the global best position ${\bf x}^{\rm G}=\max_i {\bf x}^{\rm P}[i]$\;
 \While{the stopping criterion is not met}{
  \While{$i<I$}{
 Update ${\bf u}^{\rm Pin}{[i]}$ and ${\bf x}_i^{\rm Pin}[i]$ by \eqref{update:u} and \eqref{update:x}, respectively\;
 Adjust ${\bf x}^{\rm Pin}{[i]}$ by \eqref{ad:1} and \eqref{ad:2}\;
 Set $\widetilde{\bf w}_{0}={\bf 0}_2$ and $\{\widetilde{\bf w}_{k}\}_{k\ne 0}$ via \eqref{w} \;
 Evaluate the fitness function value by \eqref{fitness}\;
  \If{${\cal F}({\bf x}^{\rm Pin}[i])>{\cal F}({\bf x}^{\rm P}[i])$}{{Update ${\bf x}^{\rm P}[i]\leftarrow{\bf x}^{\rm Pin}[i]$}\;}
\If{${\cal F}({\bf x}^{\rm Pin}[i])>{\cal F}({\bf x}^{\rm G})$}{{Update ${\bf x}^{\rm G}\leftarrow{\bf x}^{\rm Pin}[i]$}\;}
 }
 }
\end{algorithm}

\newtheorem{Rem}{Remark}
\begin{Rem}
We note that the feasibility module (including \eqref{ad:1} and \eqref{ad:2}) to adjust PAs to satisfy their placement constraints can be applied to other heuristic algorithms as well as deep learning-based algorithms.  
\end{Rem}

\subsection{Second Stage: Baseband Beamforming Design}

The first stage determines the {high-quality feasible}  PA positions with preliminary beamforming designs, while the second stage aims  to refine beamforming vectors to further  enhance overall system performance. For given $\{\widetilde{\bm\psi}^{\rm Pin}_{l,n}\}$, we reformulate the two problems via variable substitution  combined with first-order Taylor approximation, and subsequently solve them using the SCA method. For notational simplicity, we use ${\bf f}_k$ to represent ${\bf f}_k(\{{\bm\psi}_{l,n}^{\rm Pin}\})$ in this subsection.

\subsubsection{SSR Maximization Design}

 Problem $\rm P_1$ is given by
\begin{subequations}
	\begin{flalign}
		{{\bf P}_{\rm 1A}}: \max_{\left\{\sum\nolimits_{k=0}^K\left\|{\bf w}_k\right\|^2 \le P_{\max}  \right\}}\sum\limits_{k=1}^K R_{\rm Sec}\left(\left\{\widetilde{\bm \psi}^{\rm Pin}_{l,n},~{{\bf w}_{k}}  \right\}\right),
	\end{flalign}
\end{subequations}
which aims to maximize SSR based on the updated PA positions obtained by Algorithm 1. Problem $\rm P_{1A}$ is non-convex. We propose an efficient algorithm to solve it based on the SCA method as follows.

Define auxiliary variables $\{\alpha_k\}$ and  $\{\beta_k\}$ as 
\begin{flalign}
&\alpha_k = R_k\left(\left\{\widetilde{{\bm\psi}}_{l,n}^{\rm Pin},{\bf w}_k\right\}\right),\label{alpha1}\\
{\rm and}~&\beta_k = R^{\rm Eve}_k\left(\left\{\widetilde{\bm\psi}_{l,n}^{\rm Pin},{\bf w}_k\right\}\right).\label{beta1}
\end{flalign}
Substituting  $\{\alpha_k\}$ and  $\{\beta_k\}$ into Problem $\rm P_{1A}$, and we have an equivalent form of Problem $\rm P_{1A}$ as
\begin{subequations}
	\begin{flalign}
		&{{\bf P}_{\rm 1B}}: \max_{\left\{{\bf w}_k,\alpha_k,\beta_k  \right\}}\sum\nolimits_{k=1}^K \left(\alpha_k-\beta_k \right)\\
        {\rm s.t.}~&{\left|{\bf f}_k{\bf w}_k\right|^2} \ge \left(2^{\alpha_k}-1\right)\left({\sum\limits_{k^{\prime}=0,k^{\prime}\ne k}^K\left|{\bf f}_k{\bf w}_{k^\prime}\right|^2+\sigma^2_k}\right),\label{p1b:a}\\
        &{\left|{\bf f}_0{\bf w}_k\right|^2} \le \left(2^{\beta_k}-1\right)\left({\sum\limits_{k^{\prime}=0,k^{\prime}\ne k}^K\left|{\bf f}_0{\bf w}_{k^\prime}\right|^2+\sigma^2_0}\right),\label{p1b:b}\\
        &\alpha_k-\beta_k\ge0,\eqref{cons:1}.
	\end{flalign}
\end{subequations}

We introduce auxiliary variables $\{a_k,b_k,c_k,d_k,f_k\}$ satisfying that 
\begin{flalign}
&e^{{a}_k} \ge 2^{{\alpha}_k}-1, \label{auv1}\\
&e^{b_k} \ge {\sum\nolimits_{k^{\prime}=0,k^{\prime}\ne k}^K\left|{\bf f}_k{\bf w}_{k^\prime}\right|^2+\sigma^2_k}\label{auv2},\\
&e^{c_k} \le  2^{{\beta}_k}-1\label{auv3},\\
&e^{d_k}\le{\sum\nolimits_{k^{\prime}=0,k^{\prime}\ne k}^K\left|{\bf f}_0{\bf w}_{k^\prime}\right|^2+\sigma^2_0}\label{auv4},\\
{\rm and}~&c_k+d_k\ge f_k,\label{auv5}
\end{flalign}
and substitute \eqref{auv1}-\eqref{auv5} into  Problem $\rm P_{1B}$ to obtain  
\begin{subequations}
	\begin{flalign}
		&{{\bf P}_{\rm 1C}}: \max_{\left\{{\bf w}_k,\alpha_k,\beta_k,a_k,b_k,c_k,d_k,f_k  \right\}}\sum\nolimits_{k=1}^K \left(\alpha_k-\beta_k \right)\\
        {\rm s.t.}~&{\left|{\bf f}_k{\bf w}_k\right|^2} \ge e^{a_k+b_k},\label{p1c1}\\
        &{\left|{\bf f}_0{\bf w}_k\right|^2} \le e^{f_k},\label{p1c2}\\
        &\alpha_k-\beta_k\ge0,\label{p1c3}\\
        &\eqref{cons:1},\eqref{auv1},\eqref{auv2},\eqref{auv3},\eqref{auv4},\eqref{auv5}.\nonumber
	\end{flalign}
\end{subequations}
Problem $\rm P_{1B}$ is non-convex due to constraints \eqref{auv1}, \eqref{auv2}, \eqref{auv3}, \eqref{auv4}, \eqref{p1c1}, and \eqref{p1c2}. Nevertheless, both sides of \eqref{auv1}, \eqref{auv2}, \eqref{auv3}, \eqref{auv4}, \eqref{p1c1},  and \eqref{p1c2} are convex functions. Therefore, we can apply the first-order Gaussian approximation to \eqref{auv1}, \eqref{auv2}, \eqref{auv3}, \eqref{auv4}, \eqref{p1c1}, and \eqref{p1c2} to derive their approximate forms:
\begin{flalign}
     &e^{{\widetilde a}_k}\left(1+a_k - {\widetilde a}_k \right) \ge 2^{{\alpha}_k}-1,\label{app_form1}\\
        &e^{{\widetilde b}_k}\left(1+b_k - {\widetilde b}_k \right) \ge {\sum\limits_{k^{\prime}=0,k^{\prime}\ne k}^K\left|{\bf f}_k{\bf w}_{k^\prime}\right|^2+\sigma^2_k},\label{app_form2}\\
        &2^{{\widetilde \beta}_k}+2^{{\widetilde \beta}_k}\ln2\left(\beta_k - {\widetilde \beta}_k \right)\ge e^{c_k} + 1,\label{app_form3}\\
        &{\sum\nolimits_{k^{\prime}=0,k^{\prime}\ne k}^K\left(2{\rm Re}\left\{{\widetilde{\bf w}}_{k^\prime}^H{\bf f}_0^H{\bf f}_0{\bf w}_{k^\prime} \right\}-{\left|{\bf f}_0{\widetilde{\bf w}}_{k^\prime}\right|^2}\right)+\sigma^2_0}\nonumber\\
        &\ge e^{d_k},\label{app_form4}\\
        &2{\rm Re}\left\{{\widetilde{\bf w}}_k^H{\bf f}^H_k{\bf f}_k{\bf w}_k \right\}-{\left|{\bf f}_k{\widetilde{\bf w}}_k\right|^2} \ge e^{a_k+b_k},\label{app_form5}\\
        {\rm and}~& e^{{\widetilde f}_k}\left(1+f_k - {\widetilde f}_k \right) \ge {\left|{\bf f}_0{\bf w}_k\right|^2},\label{app_form6}
\end{flalign}
where $\{\widetilde{\bf w}_{k},{\widetilde \beta}_k,{\widetilde a}_k, {\widetilde b}_k, {\widetilde f}_k\}$ are feasible to Problem $\rm P_{1C}$. 

We then obtain an approximate yet convex formulation of Problem $\rm P_{1A}$, given as follows:
\begin{subequations}
	\begin{flalign}
		&{{\bf P}_{\rm 1D}}: \max_{\left\{{\bf w}_k,\alpha_k,\beta_k,a_k,b_k,c_k,d_k,f_k  \right\}}\sum\nolimits_{k=1}^K \left(\alpha_k-\beta_k \right)\\
        {\rm s.t.}~&\eqref{cons:1},\eqref{auv5},\eqref{p1c3},\eqref{auv1}, \eqref{auv2}, \eqref{auv3}, \eqref{auv4}, \eqref{p1c1}, \eqref{p1c2}.\nonumber
	\end{flalign}
\end{subequations}

Note that the approximation from Problem $\rm P_{1A}$  to Problem $\rm P_{2D}$ is based on the first-order Gaussian approximation. To enhance  the approximation precision, the SCA method can be employed to obtain a stationary-point solution to Problem $\rm P_{1A}$, the procedure of which is summarized in Algorithm \ref{alg1}.

\subsubsection{SEE Maximization Design} 
We can directly extend the  SSR  maximization  algorithm for SEE maximization via the following processes.

Given $\{\widetilde{\bm\psi}^{\rm Pin}_{l,n}\}$, Problem $\rm P_2$ is given by
\begin{subequations}
	\begin{flalign}
		{{\bf P}_{\rm 2A}}: \max_{\left\{\sum\nolimits_{k=0}^K\left\|{\bf w}_k\right\|^2 \le P_{\max}  \right\}} \frac{\sum\nolimits_{k=1}^K R_{\rm Sec}\left(\left\{\widetilde{\bm \psi}^{\rm Pin}_{l,n},~{{\bf w}_{k}}  \right\}\right)}{\sum\nolimits_{k^{\prime}=0}^K\left\|{\bf w}_{k^{\prime}}\right\|^2 + P_{\rm C}}.
	\end{flalign}
\end{subequations}

We introduce additional auxiliary variables $\{\lambda_k\}$ satisfying 
\begin{flalign}
\lambda_k = \frac{\alpha_k - \beta_k}{\sum\nolimits_{k^{\prime}=0}^K\left\|{\bf w}_{k^{\prime}}\right\|^2 + P_{\rm C}},
\end{flalign}
and substitute $\{\alpha_k\}$, $\{\beta_k\}$ (defined in \eqref{alpha1} and \eqref{beta1}, respectively) and $\{\lambda_k\}$ into Problem $\rm P_{2A}$ to obtain its equivalent form:
\begin{subequations}
	\begin{flalign}
		{{\bf P}_{\rm 2B}}:&\max_{{\left\{{\bf w}_k,\alpha_k,\beta_k,\lambda_k  \right\}}}  \sum\nolimits_{k=1}^K{\lambda_k} \label{p1b}\\
        {\rm s.t.} &  \alpha_k-\beta_k\ge\lambda_k\left({\sum\nolimits_{k^{\prime}=0}^K\left\|{\bf w}_{k^{\prime}}\right\|^2 + P_{\rm C}}\right),\label{p2b1}\\
        &\eqref{cons:1},\eqref{p1b:a},\eqref{p1b:b}.\nonumber
	\end{flalign}
\end{subequations}

We define auxiliary variables $\{g_k\}$ and $\{h_k\}$ satisfying that
\begin{flalign}
&e^{g_k} \ge\lambda_k,\label{auv_g}\\
{\rm and}~&e^{h_k} \ge {\sum\nolimits_{k^{\prime}=0}^K\left\|{\bf w}_{k^{\prime}}\right\|^2 + P_{\rm C}},\label{auv_h}
\end{flalign}
and substitute $\{a_k,b_k,c_k,d_k,f_k\}$ (defined in \eqref{auv1}-\eqref{auv5}, respectively), $\{g_k\}$ and $\{h_k\}$  into Problem $\rm P_{2B}$ to obtain its equivalent form:
\begin{subequations}
	\begin{flalign}
		&{{\bf P}_{\rm 2C}}: \max_{\left\{{\bf w}_k,\alpha_k,\beta_k,\lambda_k,a_k,b_k,c_k,d_k,f_k,g_k,h_k  \right\}}\sum\nolimits_{k=1}^K \lambda_k\label{p2c}\\
        {\rm s.t.}~&\alpha_k - \beta_k \ge e^{g_k+h_k},\label{p2c1}\\
        &\eqref{cons:1},\eqref{auv1},\eqref{auv2},\eqref{auv3},\eqref{auv4},\eqref{auv5},\eqref{p1c1},\eqref{p1c2},\nonumber\\
        &\eqref{p1c3},\eqref{auv_g},\eqref{auv_h}.\nonumber
	\end{flalign}
\end{subequations}
Problem $\rm P_{2C}$ is non-convex. Similarly, we apply the first-order Gaussian approximation to non-convex constraints \eqref{auv1}, \eqref{auv2}, \eqref{auv3}, \eqref{auv4}, \eqref{p1c1},  \eqref{p1c2}, \eqref{auv_g} and \eqref{auv_h} to obtain an approximation form of Problem $\rm P_{2C}$:
\begin{subequations}
	\begin{flalign}
		{{\bf P}_{\rm 2D}}:&\max_{{\left\{{\bf w}_k,\alpha_k,\beta_k,a_k,b_k,c_k,d_k,f_k,\lambda_k,g_k,h_k  \right\}}}  \sum\nolimits_{k=1}^K{\lambda_k}  \label{p2b}\\
        {\rm s.t.}~&e^{{\widetilde g}_k}\left(1+g_k - {\widetilde g}_k \right) \ge\lambda_k,\\
        &e^{{\widetilde h}_k}\left(1+h_k - {\widetilde h}_k \right) \ge {\sum\nolimits_{k^{\prime}=0}^K\left\|{\bf w}_{k^{\prime}}\right\|^2 + P_{\rm C}},\\
        &\eqref{cons:1},\eqref{auv5},\eqref{p1c3},\eqref{auv1},\eqref{auv2},\eqref{auv3},\eqref{auv4},\eqref{auv1}, \eqref{auv2}, \nonumber\\
        &\eqref{auv3}, \eqref{auv4}, \eqref{p2c1},\nonumber
	\end{flalign}
\end{subequations}
where $\{\widetilde{\bf w}_{k},{\widetilde \beta}_k,{\widetilde a}_k, {\widetilde b}_k, {\widetilde f}_k,{\widetilde g}_k,{\widetilde h}_k\}$ are feasible to Problem $\rm P_{2D}$.

By iteratively solving Problem $\rm P_{2D}$ using the SCA method, a stationary-point solution to Problem $\rm P_{2A}$ can be obtained. The overall algorithm is summarized in Algorithm \ref{alg1}.

\begin{algorithm}[!t]\label{alg1}
\caption{SCA-based algorithm for solving Problem $\rm P_{1A}/P_{2A}$.}
Initialize  ${\widetilde{\bf w}_{\rm k}}$ satisfying \eqref{cons:1}\;
Initialize $\{{\widetilde \beta}_k,{\widetilde a}_k, {\widetilde b}_k, {\widetilde f}_k\}$  and $\{{\widetilde g}_k, {\widetilde h}_k\}$\;
 \While{the stopping criterion is not met}{
 Update ${\widetilde{\bf w}_{\rm k}}$ by the solution to Problem $\rm P_{1D}$/$\rm P_{2D}$\;
 Update $\{{\widetilde \beta}_k,{\widetilde a}_k, {\widetilde b}_k, {\widetilde f}_k\}$  and $\{{\widetilde g}_k, {\widetilde h}_k\}$\;
 }
\end{algorithm}

\subsection{Computational Complexity Analysis}

This subsection analyzes the computational complexity of the proposed two-stage  algorithm.

\subsubsection{Complexity of FeaPSO}
For each particle corresponding to a candidate PA positioning result, two core computations are executed: CSI vector calculation incurs \(\mathcal{O}(KN)\) operations, and secrecy rate fitness evaluation costs \(\mathcal{O}(K^2)\) operations. The additional feasibility correction module only requires sorting and coordinate rescaling, which introduces  \(\mathcal{O}(N\log N)\) overhead. Given I swarm particles and a maximum of \(G_{\rm PSO}\) evolutionary generations, the overall complexity of the FeaPSO stage is expressed as
\(\mathcal{O}(G_{\rm PSO} I (KN + K^2 + N\log N))\).

\subsubsection{Complexity of SCA}
After acquiring fixed PA positions from the FeaPSO stage under either parallel or orthogonal dual-waveguide placement, the SCA method is adopted to solve a sequence of convex surrogate subproblems via interior-point solvers. The number of auxiliary optimization variables scales linearly with K, such that each interior-point iteration incurs a computational cost of \(\mathcal{O}(K^{3.5})\). Let \(T_{\rm SCA}\) denote the maximum number of SCA iterations until convergence. The total complexity of the beamforming optimization stage reads
\(\mathcal{O}(T_{\rm SCA} K^{3.5})\).

\subsubsection{Overall Complexity}
Synthesizing the overhead of the two cascaded stages, the full computational complexity of the proposed algorithm is given by
\begin{flalign}
\mathcal{O}\left(G_{\rm PSO} I \left(KN + K^2 + N\log N\right) + T_{\rm SCA} K^{3.5}\right).
\end{flalign}

\subsection{Comparison of Two Placement Schemes}

 For the considered system, path loss serves as a central factor. Therefore, we compare the two waveguide placement schemes in terms of path loss. We assume that for the $k$-th receiver, its $x$- and $y$-coordinates are independent and uniformly distributed within the considered square area, i.e., $x_k\sim{\mathcal U}(-D,D)$ and $y_k\sim{\mathcal U}(-D,D)$.

Since PAs are movable, we focus on the squared Euclidean distances between LUs and waveguides, which consist of the horizontal planar distance and a fixed vertical height $h$. Intuitively, a smaller Euclidean distance yields lower path loss and thus better channel conditions for LUs. In the following, we derive the expected squared distances for the parallel and orthogonal placement schemes, respectively.

For parallel placement, the squared Euclidean distance between the $k$-th LU and the $l$-th waveguide is given by
\begin{flalign}
    {d^2_{k,l}} = \left\{ \begin{array}{l}
 {{{\left( {{y_k} - \frac{{{D}}}{3}} \right)}^2} + {h^2}},~l=1\\
 {{{\left( {{y_k} + \frac{{{D}}}{3}} \right)}^2} + {h^2}},~l=2 
\end{array} \right.,
\end{flalign}
and its expectation is given by
\begin{flalign}\label{dkl}
    \mathbb{E}\left\{{d^2_{k,l}}\right\}=\frac{{4}}{{9}}{D^2} + {h^2},~l\in\{1,2\}.
\end{flalign}

For orthogonal placement, the squared Euclidean distance is given by
\begin{flalign}
    {d^2_{k,l}} = \left\{ \begin{array}{l}
 {{{ {{x^2_k}} }} + {h^2}},~l=1, \\
 {{{{y^2_k}}} + {h^2}},~l=2, 
\end{array} \right.
\end{flalign}
and its expectation is given by
\begin{flalign}\label{dkl3}
   \mathbb{E}\left\{{d^2_{k,l}}\right\}=\frac{D^2}{3} + {h^2},~l\in\{1,2\}.
\end{flalign}
The derivation of \eqref{dkl} and \eqref{dkl3} can be found in the Appendix.

Comparing \eqref{dkl} and \eqref{dkl3}, orthogonal placement yields a smaller expected squared LU–waveguide Euclidean distance than its parallel counterpart, indicating slightly better average path-loss conditions for LUs under orthogonal placement. Note that this metric serves only as a rough indicator and does not indicate the universal superiority of orthogonal placement.

Moreover, PLS is achieved by exploiting the channel diversity between LUs and Eve. For LoS transmission, such diversity is governed by the relative positions of PAs and LUs/Eve. In general, guaranteeing information secrecy becomes challenging when Eve is located close to LUs. Fortunately, movable PAs, especially under orthogonal placement, can enhance this channel diversity and improve the PLS performance.

\section{Simulation Results}

\begin{table}[t]
\centering
\caption{Simulation Parameters.}
\begin{tabular}{c|c }
\hline
{\bf Parameter}               & {\bf Value}               \\ \hline
\hline
Number of PAs per waveguide & $N=2$ \\ 
\hline
Side length of the square serving area  & $2D=20$ m \\ \hline
Height of waveguides  & $h=3$ m \\ \hline
Carrier frequency & $f_c=6$ GHz \\ \hline
Minimum guard distance  & $\Delta=\lambda/2$ \\ \hline
Effective refractive index     & $n_{\rm neff}=1.4$ \\ \hline
Noise power & $\sigma_{\rm B}^2=\sigma_{\rm E}^2=-70$ dBm \\ \hline
Power budget & $P_{\max}=30$ dBm\\
\hline
Circuit power &  $P_{\rm C}=20$ dBm\\
\hline
\end{tabular}
\label{Simulation setting}
\end{table}

This section presents simulation results to evaluate the PLS performance of the dual-waveguide PA system. Default system parameters follow the setup given in \cite{PANOMA}. We set the number of PAs per waveguide to \(N=2\) and the number of LUs to \(K=2\). The side length of the service area square is defined as 2$D$, where \(D=10\) m. The positions of LUs and Eve are randomly sampled within this square region. The transmit power budget is set to \(P_{\max}=30\) dBm, and the circuit power consumption is \(P_{\rm C}=20\) dBm. The waveguides are mounted at a height of \(h=3\) m. The noise power at LUs and Eve satisfies \(\sigma_{\rm B}^2=\sigma_{\rm E}^2=-70\) dBm. The carrier frequency, effective refractive index of the dielectric waveguide, and minimum inter-PA guard distance are configured as \(f_c=6\) GHz, \(n_{\rm neff}=1.4\), and \(\Delta = \lambda/2\), respectively. For comparison, we also consider a fixed-position antenna system located at coordinate \([0,0,h]\), which is equipped with $N$ antennas separated by spacing \(\Delta\). This baseline adopts the same optimized digital baseband beamforming and AN framework, solved using the SCA algorithm. We adopt the in-waveguide phase shifts as  the default channel model, unless specified otherwise.

The initialization settings of the proposed two-stage algorithm are specified as follows. In first-stage FeaPSO, we set $\kappa=0.8$, $v_1=v_2=1.5$, and $I=5000$. For parallel placement, ${x}^{\rm Pin}_{l,n}{[i]}$ is initialized following an uniform distribution ${\mathcal U} (\mathrm{mean}(x_k)-1,\mathrm{mean}(x_k)+1)$. For orthogonal placement, {${y}^{\rm Pin}_{1,n}{[i]}$ and ${x}^{\rm Pin}_{2,n}{[i]}$ are initialized following ${\mathcal U}(\mathrm{mean}(y_k)-1,\mathrm{mean}(y_k)+1)$ and ${\mathcal U}(\mathrm{mean}(x_k)-1,\mathrm{mean}(x_k)+1)$, respectively.} Furthermore, ${\bf u}^{\rm Pin}{\left[i\right]}$ is initialized following ${\mathcal U}(-1.5,1.5)$. In second-stage SCA, we adopt the initialization algorithm presented in \cite{bckpls1}. {The parallel and orthogonal placement schemes are compared under identical system configurations as well as identical algorithm initialization settings.}

\subsection{Algorithm Effectiveness}

\begin{table*}[t]
\centering
\caption{{Algorithm effectiveness for Parallel Placement.}}
\begin{tabular}{c|c|c|c||c|c}
\hline
{{Problem}}&   {{Metric}} & {FeaPSO}& {SCA} & {ES-0.4} & {SCA}\\
\hline
\hline
\multirow{3}{*}{${\bf P}_1$ [bit/s/Hz]} 
 &   $\sum_k R_k$&23.16 $\pm$ 1.43 &23.88 $\pm$ 1.49& 22.43 $\pm$ 1.10 & 23.55 $\pm$ 1.29 \\
 \cline{2-6}
& $\sum_k  R^{\rm Eve}_k$ &2.16 $\pm$ 0.26 &2.03 $\pm$ 0.11& 2.24 $\pm$ 0.31& 2.06 $\pm$ 0.09 \\
\cline{2-6}
 & $\sum_k  R^{\rm Sec}_k$ &20.99 $\pm$ 1.42 &21.84 $\pm$ 1.49& 20.19 $\pm$ 1.07& 21.49 $\pm$ 1.29 \\

\hline
\hline
{${\bf P}_2$ [bit/J/Hz]} 
 & SEE & 14.00 $\pm$ 0.95 &25.47 $\pm$ 2.44 & 13.46 $\pm$ 0.72&  24.86 $\pm$ 2.10\\
\hline
\end{tabular}
\label{table:ssr:p}
\end{table*}

\begin{table*}[t]
\centering
\caption{{Algorithm effectiveness for Orthogonal Placement.}}
\begin{tabular}{c|c|c|c||c|c}
\hline
{{Problem}}&   {{Metric}} & {FeaPSO}& {SCA} & {ES-0.4} & {SCA}\\
\hline
\hline
\multirow{3}{*}{${\bf P}_1$ [bit/s/Hz]} 
 &   $\sum_k R_k$&23.43 $\pm$ 1.35&24.12 $\pm$ 1.41& 22.56 $\pm$ 1.09& 23.74 $\pm$ 1.20\\
 \cline{2-6}
& $\sum_k R^{\rm Eve}_k$ &2.17 $\pm$ 0.23 &2.04 $\pm$ 0.07& 2.24 $\pm$ 0.31 & 2.06 $\pm$ 0.09 \\
\cline{2-6}
 & $\sum_k R^{\rm Sec}_k$ &21.26 $\pm$ 1.33 &22.08 $\pm$ 1.40& 20.33 $\pm$ 1.06 & 21.68 $\pm$ 1.19 \\

\hline
\hline
{${\bf P}_2$ [bit/J/Hz]} 
 & SEE & 14.17 $\pm$ 0.89 &25.85 $\pm$ 2.31 & 13.55 $\pm$ 0.71 &  25.17 $\pm$ 2.44\\
\hline
\end{tabular}
\label{table:ssr:o}
\end{table*}

\begin{figure}[t]
    \centering
    \includegraphics[width=0.48\textwidth]{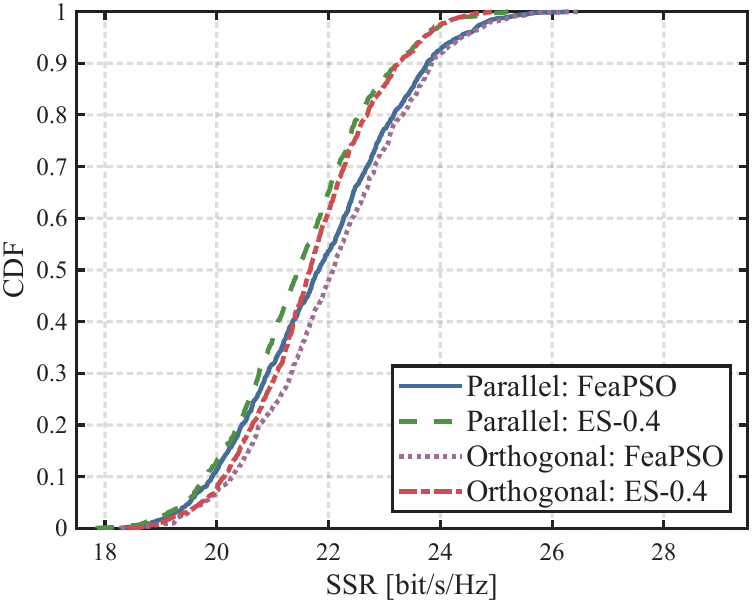}
    \caption{ The CDF comparison of SSR under different schemes.}
    \label{fig:CDF_SSR}
\end{figure}

\begin{figure}[t]
    \centering
    \includegraphics[width=0.48\textwidth]{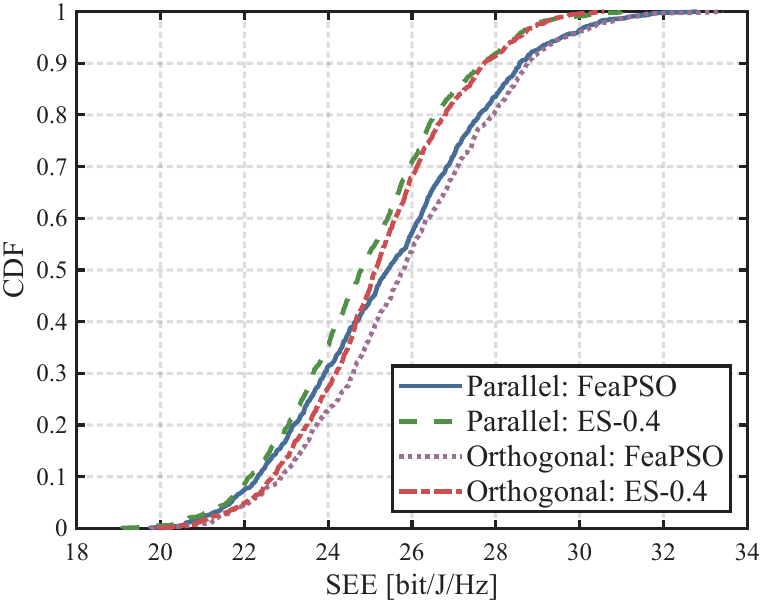}
    \caption{ The CDF comparison of SEE under different schemes.}
    \label{fig:CDF_SEE}
\end{figure}


In Tables \ref{table:ssr:p} and \ref{table:ssr:o}, we evaluate the proposed algorithm in addressing the considered problems under the two placement schemes. For comparison, we adopt the exhaustive search method (with the feasibility module) as the benchmark for the first-stage FeaPSO algorithm. {All results are averaged over $1,000$ random position realizations of LUs and Eves and are presented as means \(\pm\) sample standard deviations.} As seen, the proposed algorithm yields very close performance in comparison with the exhaustive search method\footnote{{It is worth noting that the exhaustive search benchmark  only evaluates FeaPSO’s solution quality, rather than proving its global optimality.}
} with a search spacing of $0.4$, denoted as ES-0.4. But ES-0.4 incurs $20\%$ more computational time\footnote{The computational time of the exhaustive search method is dependent on its search spacing. A smaller spacing yields better performance but incurs greater computational overhead.} than FeaPSO. These results validate the effectiveness of our proposed algorithm. In Table \ref{table:ssr:p}, we present the results for sum rate, i.e., $\sum_k R_k$, sum information leakage rate, i.e., $\sum_k  R^{\rm Eve}_k$, and SSR, i.e., $\sum_k  R^{\rm Sec}_k$. It is observed that FeaPSO can achieve a roughly {high-quality} result, while the SCA-based algorithm further refines the beamforming vectors to both enhance the achievable sum rate and reduce the sum information leakage rate, thereby improving the SSR. In Table \ref{table:ssr:o}, we present the results for SEE. It can be seen that the second stage can significantly improve SEE, which is distinct from the SSR maximization case. The reason is that in the first stage, the power budget is fully utilized as shown in \eqref{w}. Moreover, the orthogonal placement scheme delivers a slight gain in SSR and SEE over the parallel counterpart, which is consistent with the analysis in Section IV-D.

We also present the cumulative distribution function (CDF) of the achieved SSR and SEE across the $1,000$ realizations, as illustrated in Figures \ref{fig:CDF_SSR} and \ref{fig:CDF_SEE}. It can be seen that the two placement schemes yield very close performance in terms of both SSR and SEE, while orthogonal placement slightly outperforms parallel placement. This is because while the two schemes result in similar average LU/Eve-waveguide distances, orthogonal placement can introduce a new DoF, i.e., spatial arrangement of waveguides, to diversify the channel conditions between LUs and Eve. The channel diversity not only mitigates information leakage but also enhances interference management. {Furthermore, by combining the results in Tables \ref {table:ssr:p} and \ref {table:ssr:o} with those in Figures \ref {fig:CDF_SSR} and \ref {fig:CDF_SEE}, we observe that the standard deviations under both placement schemes are quite similar.}

\subsection{Impact of Key System Parameters}

\begin{figure}[t]
    \centering
    \includegraphics[width=0.48\textwidth]{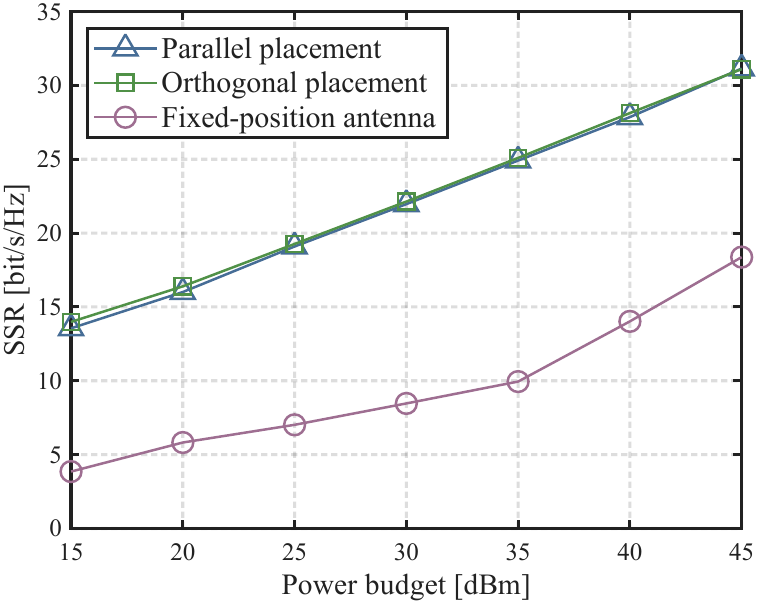}
    \caption{SSR versus power budget.}
    \label{fig:SSR_Pmax}
\end{figure}

\begin{figure}[t]
    \centering
    \includegraphics[width=0.48\textwidth]{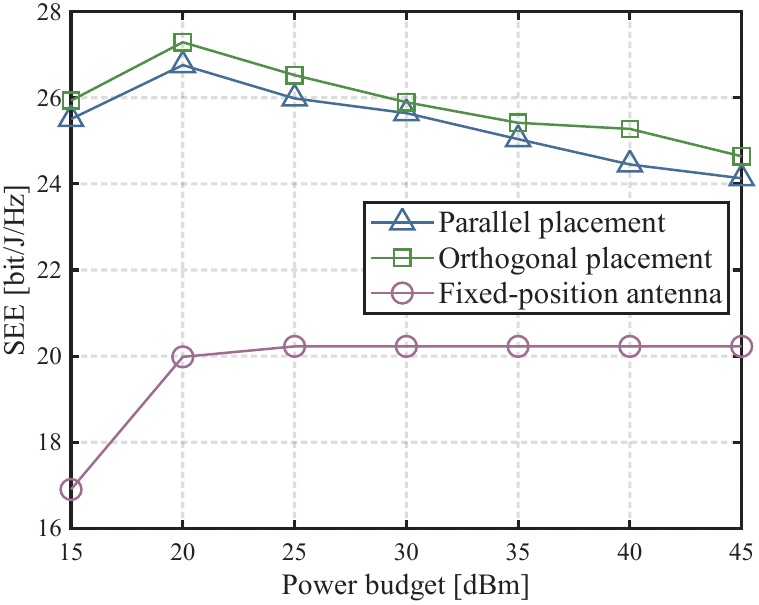}
    \caption{SEE versus power budget.}
    \label{fig:SEE_Pmax}
\end{figure}

Figure \ref{fig:SSR_Pmax} depicts the achievable SSR versus the power budget $P_{\max}$. As observed, the achievable SSR increases monotonically with $P_{\max}$ for both PAs and fixed-position antennas. The two placement schemes yield comparable SSR performance, and both clearly outperform fixed-position antennas. The reason is that movable PAs can not only reduce the transmission distance for LUs  but also actively enhance the channel diversity between LUs and Eve.  Furthermore, we use Figure \ref{fig:SEE_Pmax} to show the achievable SEE versus the power budget. As observed, SEE first increases and then slightly decreases with the increment of $P_{\max}$ for PAs. This trend can be attributed to the fact that the first stage, which fully utilizes the power budget, may introduce losses, and the losses grow slightly with $P_{\max}$.  Besides, there are minor fluctuations in the curves corresponding to PAs. Nevertheless, the PA system still achieve higher SEE compared with its fixed-position counterpart.


\begin{figure}[t]
    \centering
    \includegraphics[width=0.48\textwidth]{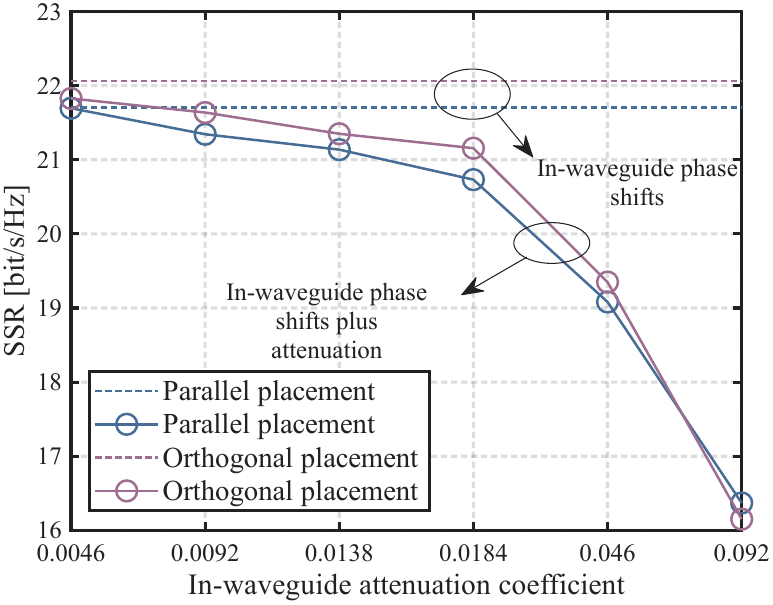}
    \caption{SSR versus the attenuation coefficient.}
    \label{fig:attenuation}
\end{figure}

\begin{figure}[t]
    \centering
    \includegraphics[width=0.48\textwidth]{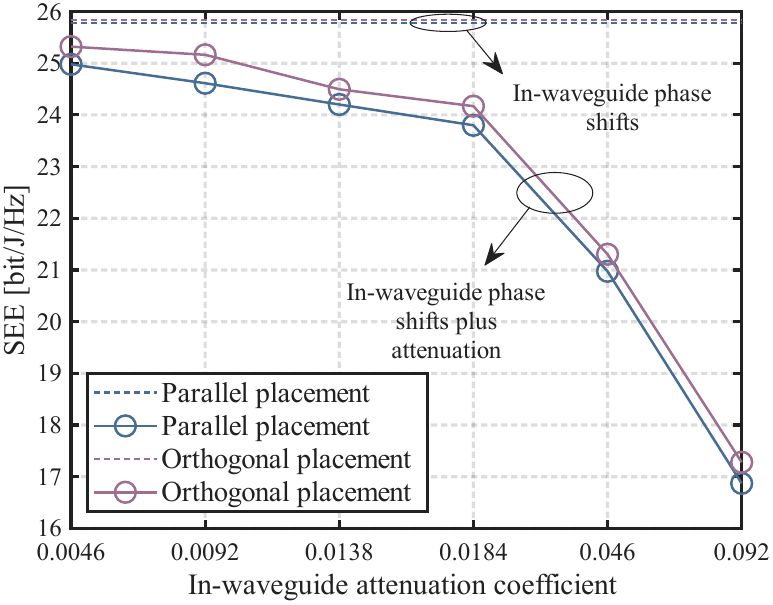}
    \caption{SEE versus the attenuation coefficient.}
    \label{fig:SEE_attenuation}
\end{figure}

Figure \ref{fig:attenuation} illustrates the impacts of the attenuation coefficient $\zeta$ (cf. \eqref{ac}) on the achievable SSR with $D$ held constant. It can be observed that as the attenuation coefficient increases, the performance gap between the two channel models, i.e, in-waveguide phase shifts and in-waveguide phase shifts and attenuation, widens progressively. The underlying reason is that in-waveguide attenuation constitutes an additional form of path loss during signal propagation from the feed point to each PA. Notably, this trend also holds for scenarios where  the attenuation coefficient is kept constant while increasing $D$. Therefore, in-waveguide attenuation can be neglected only when both $\zeta$ and $D$ take small values. Moreover, it is interesting to see that the attenuation coefficient also has impacts on the placement schemes. Therefore, the exact channel model for PAs remains a critical research challenge. Furthermore, we use Figure \ref{fig:SEE_attenuation} to show the impacts of the attenuation coefficient on SEE. We can also find that the increase in the attenuation coefficient degrades SEE for both placement schemes.



\section{Conclusion}

This paper has studied the dual-waveguide PA-enabled PLS. We proposed two placement schemes, i.e., parallel placement and orthogonal placement. Meanwhile, we have considered two channel models, i.e., in-waveguide phase shifts and in-waveguide phase shifts and attenuation. We have formulated the SSR and SEE maximization problems. To solve the problems, we have proposed a two-stage algorithm, where the first-stage FeaPSO yielded feasible PA positions and the second-stage SCA refines the beamforming vectors. Simulation results have validated the effectiveness of the proposed algorithm in the comparison with the exhaustive search method. We have derived the insights to verify the effectiveness of the orthogonal waveguide.

We conclude this paper with the following future directions: 1) Due to the deeply coupled variables, optimizing  PAs requires the development of new optimization techniques and deep learning methods. 2) The placement schemes of waveguides remain an open issue. 3) To integrate the PAs with existing flexible-antenna technologies shall provide a combinatorial gain. {4) PA adjustment latency necessitates the quantitative models to characterize the effective (secrecy) throughput of PA systems.}


\appendix
    
    \subsection{Derivation of \eqref{dkl} and \eqref{dkl3}.}

\begin{figure*}[t]
\begin{flalign}
    \mathbb{E}\left\{\left(X + A\right)^2\right\} &= \int_{{D_1} }^{{D_2} } {{{\left( {x + A} \right)}^2}}  \frac{1}{{{D_2} - {D_1}}}dx = \frac{1}{{{D_2} - {D_1}}}\left[ {\frac{{{{\left( {x + A} \right)}^3}}}{3}} \right]_{{D_1} }^{{D_2} }= \frac{{\left( {{D_1} + A} \right){}^2 + \left( {{D_1} + A} \right)\left( {{D_2} + A} \right) + {{\left( {{D_2} + A} \right)}^2}}}{3}\label{EX+A}
\end{flalign}
\hrule
\end{figure*}

For a random variable $X\sim{\mathcal U}(D_1,D_2)$, we derive the expectation of $((X+A)^2+B)$, where $D_1$, $D_2$, $A$ and $B$ are constants.

We compute $\mathbb{E}\{(X + A)^2\}$ given by \eqref{EX+A}. Then, we obtain that 
\begin{flalign}
&\mathbb{E}\left\{\left(X + A\right)^2+B\right\}= \\
&\frac{{\left( {{D_1} + A} \right){}^2 + \left( {{D_1} + A} \right)\left( {{D_2} + A} \right) + {{\left( {{D_2} + A} \right)}^2}}}{3} + B.   \nonumber
\end{flalign}

\end{document}